\documentclass{article}
\usepackage{amssymb}
\usepackage{epsfig}

\newcommand{\be}{\begin{equation}}
\newcommand{\ee}{\end{equation}}
\newcommand{\bea}{\begin{eqnarray}}
\newcommand{\eea}{\end{eqnarray}}

\title{Integrable Vortex Dynamics in Anisotropic Planar Spin Liquid Model}
\author{Zeynep Nilhan Gurkan\footnote{email: nilhangurkan@iyte.edu.tr} \\  and \\ Oktay\
Pashaev\footnote{email: oktaypashaev@iyte.edu.tr ; fax: +90 232 750 7509 (Correspondence Author)}\\
\\
 Department of Mathematics\\     Izmir Institute of Technology  \\
Gulbahce koyu, Urla, Izmir,\\35430, Turkey\\}

\begin{document}
\maketitle

\begin{abstract}
The problem of magnetic vortex dynamics in an anisotropic spin
liquid model is considered. For incompressible flow the model
admits reduction to saturating Bogomolny inequality analytic
projections of spin variables, subject the linear holomorphic
Schr\"odinger equation. It allows us to construct N vortex
configurations in terms of the complex Hermite polynomials. Using
complex Galilean boost transformations, the interaction of the
vortices and the vortex chain lattices (vortex crystals) are
studied. By the complexified Cole-Hopf transformation, integrable
N vortex dynamics is described  by  the holomorphic Burgers
equation. Mapping of the point vortex problem to N-particle
problem, the complexified Calogero-Moser system, showing its
integrability and the Hamiltonian structure, is given.

\end{abstract}

\section{Introduction}
  As is well known the main ideas of Descartes vortex theory are
represented in "Discours de la methode" (1637) and in a capital
work "Principia Philosophiae" (1644). The Cartesian cosmology is
based on primordial chaos, which by motion according to fixed laws
is ordering to cosmos \cite{Kozlov}. According to Descartes the
Universe is filled by thin all-penetrable fluid (similar to the
ether) which is in a permanent rotational motion. The term vortex
itself, "tourbillon", is coming from comparison with turbulent
motion of a river. Besides Descartes, the vortex models of gravity
were proposed by Bernoulli, Stokes and Huygens. But soon they were
displaced  by the Newton's gravity theory for a long time. Only in
the middle of XIX century the interest to vortex theory revives
with works of Helmholtz (1821-1894) \cite{Helmholtz}, Thompson
(Lord Kelvin) \cite{Kelvin}, and Kirchhoff (1824-1887)
\cite{Kirchoff} on the vortex motion of an ideal fluid. The
mathematical description of processes related with the motion of
vortex in a liquid is starting from Helmholtz's paper "Uber
Integrale der Hydrodynamischen Gleichungen Welche den
Wirbelbewegungen Entsprechen" (1858) in which he
 formulated his famous theorem on conservation of vorticity in the rotational motion of a fluid.
He also notices an analogy between the fluid motion and the
magnetic action of electric fields. General equations of motion
for N point vortices (Kirchhoff's equations) have been introduced
by Kirchhoff in his lectures in mathematical physics
\cite{Kirchoff}. He derived the corresponding Hamiltonian form of
equations and found all possible integrals of the motion. In
contrast with Newton's equation for $N$ point masses, having the
second order, the Kirchhoff equations are the first order of
vortex coordinates. Walter Gr\"obli in his thesis "Specille
Probleme uber die Bewegung Geradliniger Parallerer Wirbelfaden" in
1877 \cite{Grobli} analyzed the integrable  problem of motion of
three vortices in the plane. He obtained the system of three
nonlinear equations possessing two integrals of motion and
allowing to get explicit quadrature.  He also considered
particular case of the problem of four vortices under condition of
symmetry axis and more general problem of $2N$ vortices with N
symmetry axes.

 The interest to vortex theory increases with Kelvin's vortex theory of atoms
"On Vortex Atoms" , \cite{Kelvin}. But soon his model was
dismissed by quantum mechanical model of atoms \cite{Lomonaco}.
Kelvin also posed the problem of stability under the stationary
rotation of the system of N point vortices located at N polygon
vertices. He noticed that the problem is similar to the problem of
stability for the system of equal magnets floating in external
magnetic field. Experiments with floating magnets performed by
Mayer, leaded him to the conclusion that for number of vortices
(magnets) exceeding 5, the rotating polygon becomes unstable (in
fact the case $N = 6$ is stable). The linear stability of polygon
has been studied by J.J. Thompson. He found that for $n \le 6$ the
linear stability takes place, while for $n \geqslant 8$ is not.
Stability of  the case $n = 7$ needs nonlinear analysis and after
several failed attempts has been proved only recently.
Non-integrability of the four vortex problem in plane, indicating
on chaotization of the vortex motion, was found recently
\cite{Ziglin}.

Many problems involving interfacial motion \cite{Lushnikov} can be
cast in the form of vortex sheet dynamics \cite{Saffman}. The
discovery of coherent structures in turbulence, increases
expectations that the study of vortices will lead to models and an
understanding of turbulent flow, one of the great unsolved
problems of classical physics. Vortex dynamics is a natural
paradigm for the field of chaotic motion and modern dynamical
system theory \cite{Poincare}. The theory of line vortices and
vortex rings is a part of modern theory of liquid Helium II.
Interaction of vortex structures essentially influences on
processes in atmosphere and the ocean. In techniques, complete
understanding of friction to the motion, noise generation and
shock waves, is impossible without clear theory of the vortex
motion. In a wide and important class of motions of ideal inviscid
fluid \cite{Melvan}, the vortex dynamics provides physically
profound examples of nonlinear Hamiltonian systems of infinite
dimensions, attracting much interest in relation with chaotic
phenomena in dynamical systems \cite{Borisov}.

Modern applications of vortices extend from liquid crystals and
ferromagnets \cite{Kuratsuji}, \cite{Komineas} to superfluids
\cite{Ho}, \cite{Mermin} and from non-equilibrium patterns to
Quantum Hall effect \cite{Ezawa} and cosmic strings \cite{Thess},
\cite{Correa}, \cite{Pismen}. They can play essential role in the
loop gravity and quantum spacetime, leading to formation of the
Cantorian Space \cite{Naschie1}- fractalization of the microspace
geometry, which has been discussed recently in the context of the
fractional Quantum Hall Effect \cite{Naschie2}, \cite{Cruz} and
Cosmic Strings \cite{Naschie3}.

Recently it was shown that in  ferromagnetic nanomagnets, for
particles of various shapes (strips and rings), the switching
process involves domain walls which are composite objects made of
two or more vortices and edge defects with integer or fractional
winding numbers \cite{Oleg}. It may provide a basic model of
complex magnetization dynamics in nanomagnets by reducing it to
the creation, propagation, and annihilation of a few planar
topological defects \cite{Naschie4}. The planar vortices represent
important class of magnetic systems admitting topologically
nontrivial solutions, the dynamics of which is one of the most
intriguing questions. Due to nonlinearity in general it is very
difficult task to construct exact analytical solution even for one
vortex. This is why simplified models admitting exact treatment
are important tools to study more realistic situation. In the
present paper we consider the simple model of anisotropic planar
spin liquid, admitting construction of exact N-vortex and N-
vortex lattice solutions with integrable dynamics for arbitrary N.

\section{Topological Spin Liquid Model}

Inspired by superfluid motion at $T = 0$ in $He^3-A$, the model
introduces the normal velocity of the fermionic liquid as an
additional hydrodynamical variable, describing the background
fermionic vacuum. The magnetic fluid or spin liquid is
characterized by the local magnetization field $\vec{S}(x,y,t)$
subject to the modified Heisenberg model in the moving frame with
velocity $\vec{v}(x,y,t)$. Moreover, for planar magnetic systems
the existence of topologically nontrivial vortex configurations
requires the fluid to be rotational with non-vanishing vorticity
function \cite{Martina}. The system is
\begin{equation}
\vec{S}_{t}+\upsilon_{1}\partial_{1}\vec{S}-\upsilon_{2}\partial_{2}\vec{S}=
\vec{S}\times(\partial_{1}^{2}-\partial_{2}^{2})\cdot
\vec{S}\label{maintopmag1}
\end{equation}
\begin{equation}
\partial_{1}\upsilon_{2}-\partial_{2}\upsilon_{1}=
2\vec{S}(\partial_{1}\vec{S}\times\partial_{2}\cdot
\vec{S})\label{maintopmag2}
\end{equation}
where $\vec{S}^2(x,y,t)=1$ is classical spin field,
$\vec{v}=(v_1,v_2)$ is velocity field. The first equation is
anisotropic in $x$ and $y$ directions, the second equation,
requiring that the fluid vorticity in the plane is proportional to
the corresponding magnetic topological current component, is known
in the theory of superfluid $He^3$ as the Mermin-Ho relation
\cite{Ho}, \cite{Mermin}. This system in general should be
supplied with the continuity equation $\rho_t + \nabla \cdot (\rho
\vec{v}) = 0$ for the density $\rho (x,y,t)$. For incompressible
flow $\rho_t = 0$ the last equation simplifies to $\nabla \cdot
 \vec{v} = 0$ and allows one exclude $\rho$ from
consideration. Moreover, for the fluid flow constrained by the
incompressibility condition
\begin{equation}
\partial_{1}\upsilon_{1}+\partial_{2}\upsilon_{2}= 0 ,\label{incompres}
\end{equation}
the conservation law
\begin{equation}
\partial_{t}J_{0}+\partial_{2}J_{2}-\partial_{1}J_{1}=0
\end{equation}
 holds \cite{Martina4}, where
 \begin{eqnarray}
 J_{0}&=&(\partial_{1}\vec{
S})^2+(\partial_{2}\vec{S})^2,\label{energy}
\\
J_{1}&=&-2\partial_{1}\vec{S}\cdot\vec{S} \times
(\partial_{1}^{2}-\partial_{2}^{2})\vec{S} +
v_{1}J_{0}+2\vec{S}\cdot(\partial_{1}\vec{S} \times
\partial_{2}^{2}\vec{S}
-\partial_{1}\partial_{2}\vec{S}\times\partial_{2}
\vec{S})\nonumber \\
J_{2}&=& 2\partial_{2}\vec{S}\cdot\vec{S} \times
(\partial_{1}^{2}-\partial_{2}^{2})\vec{S} + v_{2}J_{0} -2
\vec{S}\cdot(\partial_{1}^{2}\vec{S} \times
\partial_{1}\partial_{2}\vec{S} -\partial_{1}\vec{S}
\times\partial_{2}\vec{S}).\nonumber
\end{eqnarray}

Due to this, for the incompressible flow (\ref{incompres}) the
energy functional
\begin{equation}E = \int\int J_0 d^2 x
=\int\int\{(\partial_{1}\vec{S})^{2}+(\partial_{2}\vec{S})^{2}\}
d^{2}x
\end{equation} is the conserved quantity.  The topological
charge or the winding number of a spin configuration is defined as
\begin{equation}
Q=\frac{1}{4\pi}\int\int\vec{S}\cdot(\partial_{1}\vec{S}\times
\partial_{2}\vec{S})d^{2}x \label{topologicalcharge}
\end{equation}
and it is also the conserved quantity. These two quantities are
related by the Bogomolny type inequality
\begin{equation}E \geq 8\pi|Q|.\end{equation}
It follows from the evident one
\begin{equation}
\int\int(\partial_{i}\vec{S}\pm\epsilon_{ij}\vec{S}\times
\partial_{j}\vec{S})^{2}d^{2}x\geq 0
\end{equation}
and is saturated by time dependent spin configurations satisfying
the self-duality equations
\begin{equation}
\partial_{i}\vec{S}\pm\epsilon_{ij}\vec{S}\times
\partial_{j}\vec{S}= 0\label{selfdual}
\end{equation}
If the spin vector phase space, the 2D sphere, we consider as the
Riemann sphere for a complex plane, we can project points on this
sphere to that plane by the stereographic projections
\begin{equation}
S_+=S_{1}+ iS_{2}=\frac{2\zeta}{1+|\zeta|^{2}} ,\,\,\,\,\,\,\
S_{3}=\frac{1-|\zeta|^{2}}{1+|\zeta|^{2}},\label{spanal3}
\end{equation} where
$\zeta(x,y,t)$ is a complex valued function. Now we rewrite the
self-duality equations (\ref{selfdual}) in stereographic
projection form. By the complex derivatives, for the first sign in
(\ref{selfdual}) we have the analyticity or the holomorphicity
condition: \be \zeta_{\bar{z}}(x,y,t)=0 \ee while for the second
sign we have the anti-analyticity or the anti-holomorphicity
condition: \be \zeta_{z}(x,y,t)=0 \ee where $z=x+iy$ and
$\partial_z=\frac{1}{2}(\partial_x-i\partial_y)$, $\partial_{\bar
z}=\frac{1}{2}(\partial_x+i\partial_y)$. These conditions written
in terms of the real $\Re \zeta$ and imaginary $\Im \zeta$ parts
of function $\zeta$, representing the Cauchy-Riemann equations,
describe the incompressible and irrotational fluid flow with the
velocity potential $\Re \zeta$ and the stream function $\Im \zeta$
\cite{Lavrentiev}.

\section{Anti-Holomorphic Reduction}
As we can see, analytic/anti-analytic configurations saturate
Bogomolny inequality and have minimal energy. This suggests to
find solutions of the model (\ref{maintopmag1}) with
holomorphic/anti-holomorphic stereographic projections. For this
reason we first rewrite (\ref{maintopmag1}) and
(\ref{maintopmag2}) in stereographic projection form in complex
coordinates $z$ for complex velocities $v_{\pm}= v_{1}\pm iv_{2}$
and if $\zeta$ is anti-holomorphic $\zeta_{z}=0$, then the system
is reduced to
\begin{equation}
i\zeta_{t}+iv_{+}\zeta_{\bar{z}}+2\zeta_{\bar{z}\bar{z}}-4
\frac{\zeta_{\bar{z}}^{2}\bar{\zeta}}{1+|\zeta|^{2}}=0\label{anti-holo},
\end{equation}
\begin{equation}
\partial_{z}v_{+}-\partial_{\bar{z}}v_{-}
=-8i\frac{\bar{\zeta_{z}}\zeta_{\bar{z}}}{(1+|\zeta|^{2})^{2}}
\label{compvort}.\end{equation} To be consistent, this
anti-holomorphicity constraint must be compatible with time
evolution. By direct computation we can show that for
incompressible flow (\ref{incompres}) the anti-holomorphic
constraint $\zeta_{z}=0$ is compatible with the time evolution
$\partial \zeta_{z}/ \partial t=0.$ For this flow from
(\ref{anti-holo}) and (\ref{compvort}) we have $
\zeta_{t}+F\zeta_{\bar{z}}=0$ and $F_z =0$, where $ F\equiv
v_{+}-2i\left[\ln\frac{\zeta_{\bar{z}}}{1+|\zeta|^{2}}\right
]_{\bar{z}} $.

\section{Anti-Holomorphic Ishimori Model}
The above consideration suggests to solve the incompressibility
conditions explicitly. So we consider the topological magnet model
(\ref{maintopmag1}), (\ref{maintopmag2}) with incompressibility
condition (\ref{incompres}). Equation $\vec{\nabla}\cdot
\vec{v}=0$ can be solved in terms of the stream function of the
flow, $v_{1}=\partial_{2}\psi, \,\, v_{2}=-\partial_{1}\psi$, so
that we get the so called Ishimori Model \cite{Ishimori}.
\begin{equation}
\vec{S}_{t}+\partial_2\psi\partial_1\vec{S}+\partial_1 \psi
\partial_2\vec{S}=\vec{S}\times (\partial^{2}_1\vec{S}-\partial^{2}_2\vec{S})\label{ishmod1} \end{equation}
\begin{equation}
(\partial_{1}^{2}+\partial_{2}^{2})\psi=-2\vec{S}\cdot(\partial_1\vec{S}\times\partial_2
\vec{S}).\label{ishmod2}
\end{equation}
The Ishimori model is the first example of integrable classical
spin model in 2+1 dimensions \cite{Konopel}. It was shown to be
gauge equivalent to the Davey-Stewartson equation, representing
the 2+1 dimensional generalization of the Nonlinear Schr\"odinger
equation  \cite{Lepovskiy}, \cite{Makhankov}, \cite{Pashaev}.
Though it was solved in terms of the $\bar \partial$ problem, for
description of vortices and vortex lattices we propose more simple
and elegant method. In terms of complex variables
\begin{equation}
 v_{+}=v_{1}+iv_{2}=-2i\psi_{\bar{z}} ,\,\,\,\ v_{-}=v_{1}-iv_{2}=2i\psi_{z}
\end{equation}  for incompressible flow, preserving anti holomorphicity
constraint, we have dependence $\zeta=\zeta(\bar{z},t)$ and the
model reduces to the system
\begin{equation}
i\zeta_{t}+2\psi_{\bar{z}}\zeta_{\bar{z}}
+2\zeta_{\bar{z}\bar{z}}-\frac{4\bar{\zeta}\zeta_{\bar{z}}^{2}}{1+|\zeta|^{2}}=0
\end{equation}
\begin{equation}
\psi_{z\bar{z}}=\frac{2\bar{\zeta}_{z}\zeta_{\bar{z}}}{(1+|\zeta|^{2})^{2}}
\label{vortisity}.\end{equation}
 We can rearrange the first equation as
\begin{equation}
i\zeta_{t}+2\zeta_{\bar{z}}\{\psi-2\ln(1+|\zeta|^{2})+\ln\zeta_{\bar{z}}\}_{\bar{z}}=0
\label{zeta}.\end{equation}
 If we choose $\psi=2\ln(1+|\zeta|^{2})$  then equation (\ref{vortisity})
 \begin{equation}\psi_{z\bar{z}}=
2\left[\frac{\bar{\zeta}_{z}\zeta}{{1+|\zeta|^{2}}}\right]_{\bar{z}}=2\frac{\bar{\zeta}_{z}\zeta_{\bar{z}}}{({1+|\zeta|^{2})^2}}
\label{17}\end{equation} is satisfied automatically.
 Then from (\ref{zeta}) for function $\zeta$ we have
the anti-holomorphic time dependent Schr\"odinger equation
\begin{equation}
i\zeta_{t}+2\zeta_{\bar{z}\bar{z}}=0 \label{linearschrodinger}.
\end{equation}
By complex analog of the Cole-Hopf transformation $ u=4\zeta_{\bar
z}/ \zeta=4(Log\, \zeta)_{\bar z} $ it implies complex Burgers'
equation
\begin{equation}
iu_t+uu_{\bar z}+2u_{\bar z \bar z}=0 \label{u}.
\end{equation}
In (\ref{u}) $u$ can be interpreted as the complex velocity of
effective flow with the complex potential $f(z)=Log \, \zeta^4$.
Then every zero of function $\zeta$ corresponds to the vortex
solution (pole of complex velocity) of anti-holomorphic Burgers'
equation.

\section{Magnetic Vortices as Moving Zeroes of Hermite Polynomials}
    By stereographic projection (\ref{spanal3}) at every zero of function $\zeta (\bar z_k, t) =
0$ we have
\begin{equation}
(S_{1}+ iS_{2})(\bar z_k, t)= 0,\,\,\,\, S_{3}(\bar z_k, t) = 1
\end{equation}
From another side for the polynomial $\zeta_N$ of degree $N$  at
infinity $|z| \rightarrow \infty$
\begin{equation}
(S_{1}+ iS_{2})(\bar z_k, t)= 0,\,\,\, S_{3}(\bar z_k, t) = -1
\end{equation}
It shows that every zero corresponds to the magnetic vortex
located at that zero with the spin vector $\vec{S}$ directed up,
while at infinity it is directed down (ferromagnetic type
order)(See Fig. 1). With such boundary conditions the topological
charge $Q$ in (\ref{topologicalcharge}) is an integer valued and
characterizes the number of magnetic vortices.
\begin{figure}[h]
\begin{center} \epsfig{figure=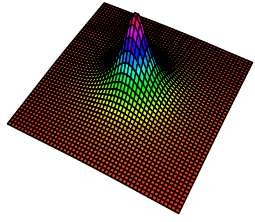,height=3.8cm,width=3.5cm}
\end{center}
\caption{$N=1$ Magnetic Vortex } \label{S3N1}
\end{figure}

If we calculate the topological charge (\ref{topologicalcharge})
for $N$ zeroes solution \begin{equation} \zeta (\bar z, t) =
\prod_{k=1}^N (\bar z - \bar z_k(t))\label{30}\end{equation} then
we find that $ Q = -N $. The above consideration shows that motion
of zeroes of time dependent Schr\"odinger equation
(\ref{linearschrodinger}) describes the motion of magnetic
vortices in the Ishimori Model. Moreover due to (\ref{u}) dynamics
of these vortices is described by the anti-holomorphic Burgers'
equation. To study motion of zeroes of (\ref{linearschrodinger})
we construct the generating function of the basic vortex solutions
by considering complex plane wave solution
\begin{equation}
\zeta(\bar{z},t)=e^{k\bar{z}+ 2ik^{2}t}.
\end{equation}
Let $x\equiv k\sqrt{\frac{2t}{i}}$, then we rewrite it as  the
generating function for the Hermite polynomials of complex
argument
\begin{equation} e^{k\bar{z}+2ik^{2}t}=
e^{-x^{2}+2(\bar{z}\sqrt{\frac{i}{8t}})x}
=\sum_{n=0}^{\infty}H_{n}\left(\bar{z}\sqrt{\frac{i}{8t}}\right)\frac{x^{n}}{n!}
\end{equation} or \begin{equation}
\zeta(\bar{z},t)=\sum_{n=0}^{\infty}\frac{k^{n}}{n!}(-2it)^{n/2}H_{n}
\left(\bar{z}\sqrt{\frac{i}{8t}}\right)
=\sum_{n=0}^{\infty}\frac{k^{n}}{n!}\Psi_{n}(\bar{z},t)
\end{equation} where at every power $k^{n}$ we have a polynomial
solution of order n:
\begin{equation}
 \Psi_{n}(\bar{z},t)=\left(\frac{2t}{i}\right)^{n/2}H_{n}\left(\bar{z}\sqrt{\frac{i}{8t}}\right).
\end{equation}
This polynomial has $n$ complex roots $\bar z_1(t),...,\bar
z_n(t)$ describing positions of vortices. For complex zeroes of
this function,  by identification \begin{equation}
\Psi_{N}(\bar{z},t) = \prod _{k=1}^{N}(\bar{z}-\bar{z}_{k}(t)) =
(-2it)^{N/2}H_{N}\left(\frac{\bar{z}}{2\sqrt{-2it}}\right)
\label{hermitecomp}
\end{equation} we find that $\bar{z}=\bar{z}_{k}(t)$ implies
$H_{N}(\frac{\bar{z}_k}{2\sqrt{-2it}})= 0 $.
 Denoting $w_k^{(N)}$, $k=1,...,n$  as zeroes of Hermite
polynomials, $H_{N}(w_{k}^{(N)})=0$ , we have the time dependence
for vortex positions
\begin{equation}
 \bar
z_{k}(t)=2w_{k}^{(N)}\sqrt{-2it}\label{hermitezero}.
\end{equation}
Due to reality of roots $w_k^{(N)}$, the form of our solution $
\bar z_{k}(t)=2w_{k}^{(N)}\sqrt{-2it}$ implies that all vortices
are located on diagonal lines of complex plane:
\begin{equation}\bar z_{k}(t)=|2w_{k}^{(N)}\sqrt{-2t}|e^{i
\pi/4}. \label{32}\end{equation}
 We note that since the time dependence includes
 square root of time variable $t$, then under time reflection, when $t$ is replaced by
 $-t$, position of vortices will rotate $\bar z_k \rightarrow e^{i\pi/2} \bar z_k$ on angle
 $\pi/2$. It means that under collision our vortices change
 velocity in orthogonal direction and from one diagonal line would
 be displaced to the orthogonal one. Moreover, sum of vortex
 positions, representing the center of mass of the system, is
 integral of motion located at the beginning of coordinates. \newpage In Figure 2 we
  show contour plot of basic vortex dynamics for four vortices.
\begin{figure*}[h]
\begin{center} \epsfig{figure=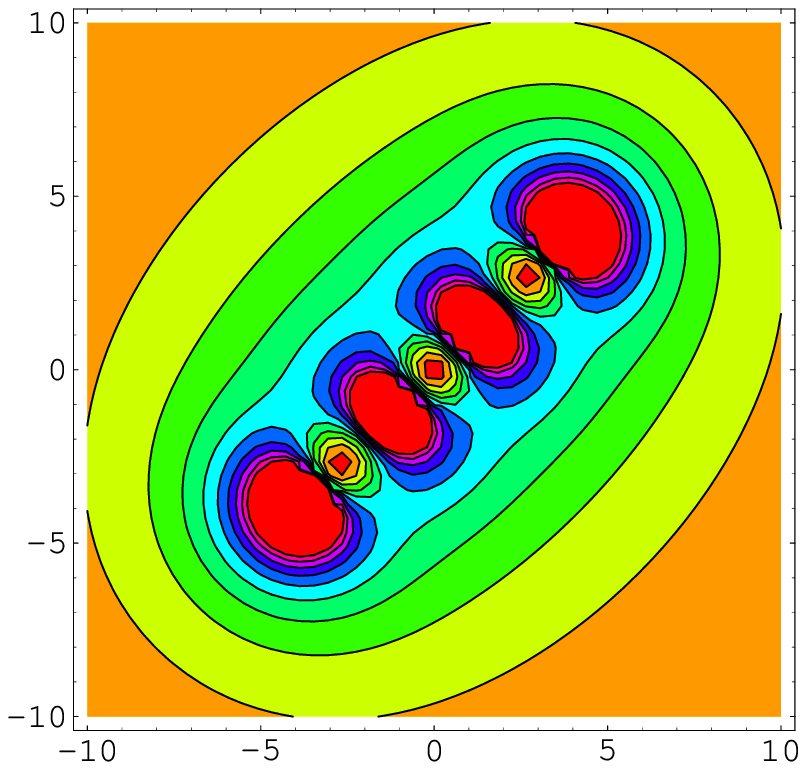,height=4cm,width=4cm}
\epsfig{figure=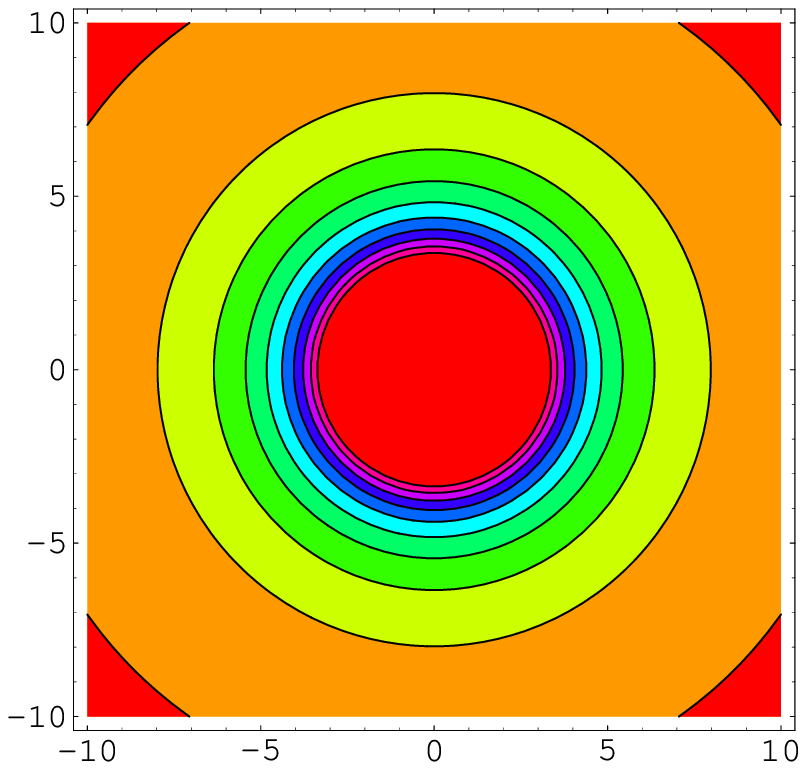,height=4cm,width=4cm}\epsfig{figure=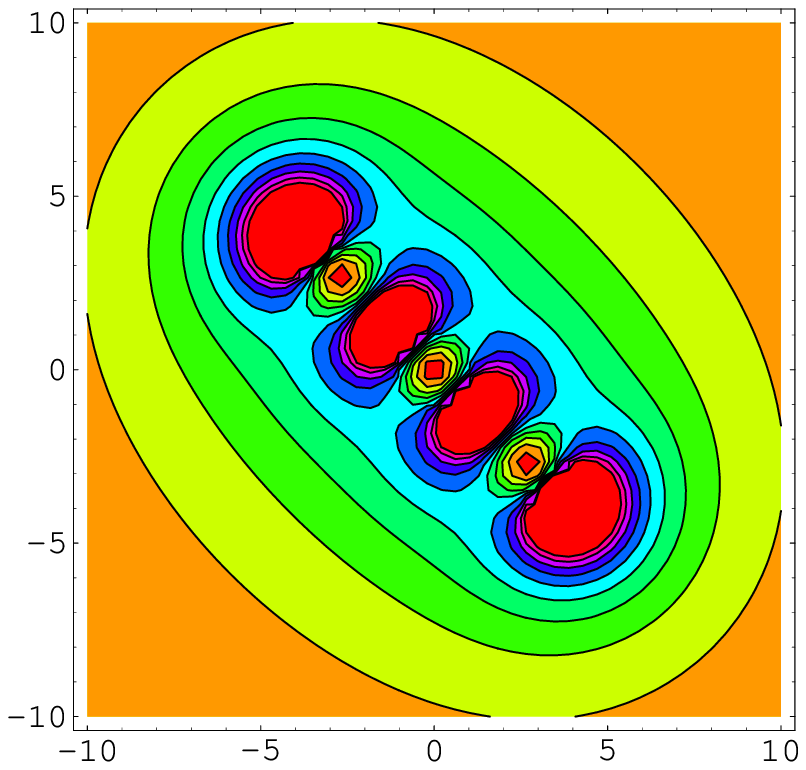,height=4cm,width=4cm}
\end{center}
\caption{$N=4$ Vortex Dynamics} \label{N4-3}
\end{figure*}
\\

In Figure 3 interaction   of two magnetic vortices in Ishimori
model is shown.
\begin{figure*}[h]
\begin{center} \epsfig{figure=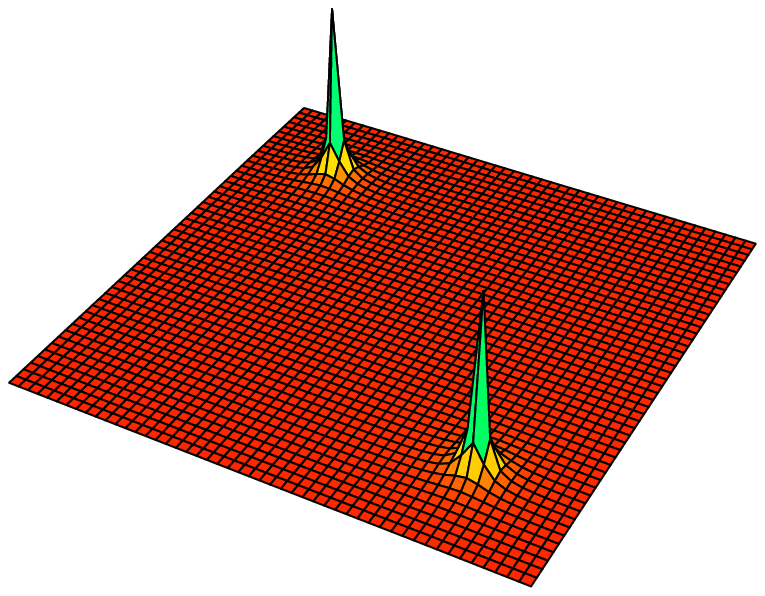,height=4cm,width=4cm}\epsfig{figure=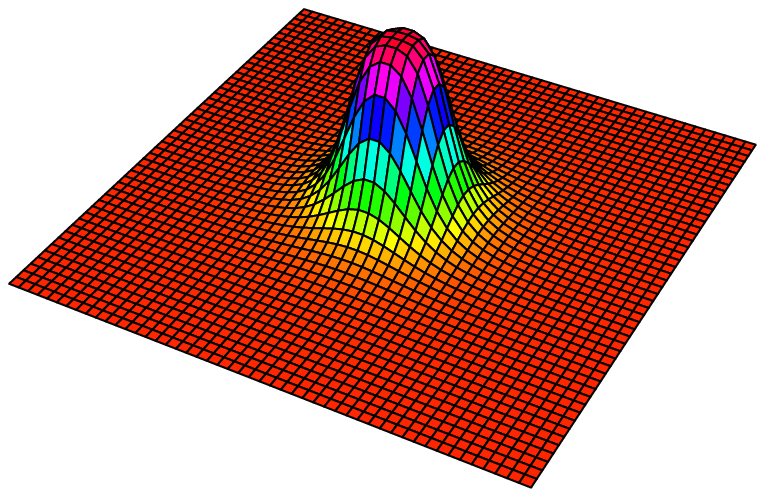,height=4cm,width=4cm}
 \epsfig{figure=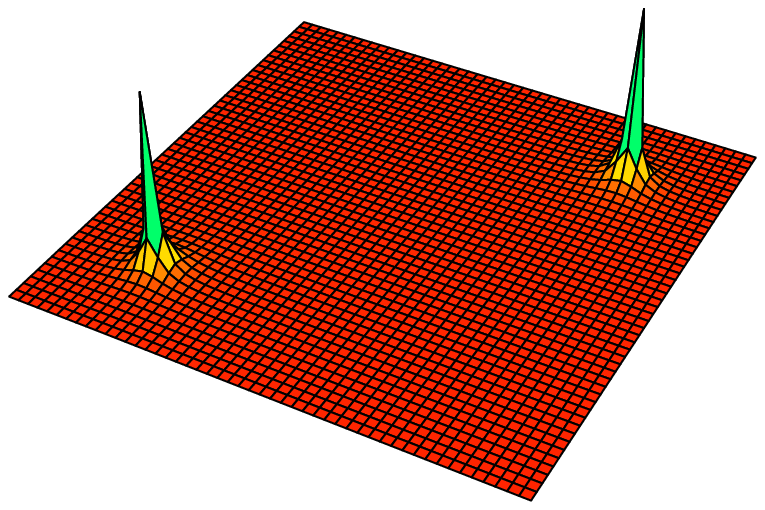,height=4cm,width=4cm}
\end{center}
\caption{$N=2$ Magnetic Vortex Dynamics} \label{S3-N2-3}
\end{figure*}

Since function $\Psi_N$ in (\ref{hermitecomp}) is solution of the
linear equation (\ref{linearschrodinger}) for any integer $N$,
then any linear combination of these functions
\begin{equation}
\Phi_N (\bar z, t) = a_N \Psi_N (\bar z, t) + a_{N-1}
\Psi_{N-1}(\bar z, t) + ... + a_0 \Psi_0(\bar z, t)
\label{superposition}\end{equation} $a_N \neq 0$ is also a
solution. This solution is determined by $N$ complex constants
$a_1, a_2, ..., a_N$, which are integrals of motion of the system.
The higher order coefficient  $a_N \neq 0$ is not essential and
could be put to one.Below we represent particular cases for $N =
3$ and $N = 4$.

1) For N = 3 we have the general solution
\begin{equation}
\Phi (\bar z, t) = (\bar z^3 + 12 \bar z i t) + a_2 (\bar z^2 + 4i
 t) + a_1 \bar z + a_0
\end{equation}
This cubic in $z$ equation has 3 complex roots $\bar z_1(t),\bar
z_2(t), \bar z_3(t)$ moving in plane according to the systems
(\ref{ishvortex1}). Instead of solving that system of differential
equations we will find roots of cubic equation according Cardano
formulas. Coefficient
$$-a_2 = \bar z_1(t) + \bar z_2(t)+ \bar z_3(t) $$
is integral of motion  having meaning of the center of mass for
three vortices. Without loss of generality we can always choose
coordinate system with the beginning at this center of mass. So we
will put $a_2 = 0$. Then our cubic equation has the reduced
Cardano form
\begin{equation}  \bar z^3 + p \bar z + q = 0
\end{equation}
 where $p = a_1 +12\bar z it$, $q = a_0 +4 a_2 it$.
Solution of this equation is
\begin{equation}
\bar z_1 = \alpha_1 + \beta_1 ,\,\,\,\,\,\, \bar z_2 = \alpha_1
\omega_1 + \beta_1 \omega_2,\,\,\,\,\,\, \bar z_3 = \alpha_1
\omega_2 + \beta_1 \omega_1
\end{equation}
 where $\alpha_1, \beta_1$ is  one
of the couple of roots
\begin{equation} \alpha =
\left(-\frac{q}{2} + \sqrt{\frac{q^2}{4}+
\frac{p^3}{27}}\right)^{1/3} ,\,\,\,\,\,\,
 \beta = \left(-\frac{q}{2} - \sqrt{\frac{q^2}{4}+
\frac{p^3}{27}}\right)^{1/3}
\end{equation}
and
$$\omega_1 = -1/2
+ i \sqrt{3}/2 = e^{i2\pi/3},\,\,\,\,\,\, \omega_2 = -1/2 - i
\sqrt{3}/2 = e^{-i2\pi/3}$$ are cubic roots of $1$. For particular
values $a_0 = a_2 = 0$ when $q = 0$ our roots become
\begin{eqnarray}
\bar z_1 &=& \sqrt{4it}(1 + e^{i \pi/3})  \\
\bar z_2 &=& \sqrt{4it}(\omega_1 + \omega_2 e^{i \pi/3}) \\
\bar z_3 &=& \sqrt{4it}(\omega_2 + \omega_1 e^{i \pi/3})
\end{eqnarray}
and coincide with particular cases (\ref{hermitezero}) when one of
the vortices, $\bar z_2 = 0$, is static at the beginning of
coordinates. In general case due to non-vanishing orbital momentum
no one of three vortices crosses beginning of coordinates.

 2) For N = 4 case we have
\begin{equation}
 \Phi (\bar z, t) = (\bar z^4 +24 \bar z^2 it + 12 (-2it)^2) + a_3(\bar z^3 +12 \bar z it) +
  a_2 (\bar z^2 +4it) + a_1 \bar z +
a_0
\end{equation}
and like in previous case we choose the center of mass at the
beginning of coordinate system, so that the coefficient
$$-a_3 = \bar z_1(t) +
\bar z_2(t)+ \bar z_3(t) + \bar z_4 (t) = 0$$ Explicit form of the
roots in this case can be done in radicals. Below we will write
them only for special case $a_1 = 0$. Solving bi-quadratic
equation $\bar z^4+ \bar z^2[a_2+24 it]+[a_0+4a_2it+12(-2it)^2]=0
$ we have four roots:
\begin{equation}
\bar z_{1,2}=\pm
\sqrt{-\left[\frac{a_2}{2}+12it\right]+\sqrt{\left(\frac{a_2}{2}+12i
t\right)^2-\left [a_0+4a_2it+12(-2it)^2\right ]}}
\end{equation}

\begin{equation}
\bar z_{3,4}=\pm \sqrt{-\left[\frac{a_2}{2}+12it\right
]-\sqrt{\left(\frac{a_2}{2}+12it\right)^2-[a_0+4a_2it+12(-2i
t)^2]}}
\end{equation} From dynamics of 3 and 4 vortices we can see
that differences appear for the vortex motion at finite times. But
for large time the behavior of vortices is similar to the
particular case (\ref{hermitezero}). Explanation of this fact can
be done for general N-vortex configuration. Indeed, if we consider
asymptotic form of the general N-vortex solution
(\ref{superposition}), when $t \rightarrow \infty$ and $\bar z
\rightarrow \infty$ such that $|z|^2/t \rightarrow const$, then we
can see that all terms of the function $\Psi_N$ for any N have the
same order. Moreover, the dominant role in (\ref{superposition})
plays the function $\Psi_N$ with highest order of N. But it is
exactly solution (\ref{hermitecomp}) which we found before. So we
proved that asymptotically our vortices will follow diagonal lines
according to the low (\ref{hermitezero}): $\bar
z_{k}(t)=2w_{k}\sqrt{-2it}$. If we calculate complex velocity
corresponding to k- vortex $u_k = d\bar z_k /dt = w_k
\sqrt{-2i/t}$ then at large times $t \rightarrow \infty$ the
velocity of the vortex is decreasing up to zero $u_k \rightarrow
0$, as the inverse square root of time $\sim 1/\sqrt{t}$.
\section{Vortex Solutions Generating Technique}
 In this section we propose a general method
allowing us to create an arbitrary number of vortices on given
background solution. In general, $N$ vortex configuration is
described by complex polynomial function degree $N$ where
coefficients are represented in terms of symmetric polynomials.
The polynomial $P_1=\bar z_1+..+\bar z_N$ is integral of motion
having meaning of the center of mass for $N$ vortices. However,
all other polynomials are not integrals of motion, only proper
combinations of these polynomials provide integrals of motion of
the system. Moreover, to find solutions $\bar z_1(t),...,\bar
z_N(t)$ in terms of these polynomials means solving algebraic
equation degree $N$, which as known to be solvable in radicals
only for $N \le 4$. This is why to add a new zero or vortex to the
system we follow another approach. In section 4 we have
constructed basic solutions in terms of Hermite polynomials of
complex argument (\ref{hermitecomp}). Adding recurrence relations
\begin{eqnarray}
H_{N+1}(x) = 2x H_N (x) - 2 N H_{N-1}(x)\\
0 = - H'_N(x) + 2 N H_{N-1}(x)
\end{eqnarray}
one can get
\begin{equation}
H_{N+1}(x) = (2x - \frac{d}{dx}) H_N (x)
\end{equation}
Recursively it leads to following operator representation of the
standard Hermite polynomials \cite{Arf}
\begin{equation} H_N(x)=(2x-\frac{d}{dx})^{N}\cdot 1
\end{equation} which implies for complex argument \begin{equation} \Psi_N(\bar
z,t)=(-2it)^{N/2}(\frac{\bar
z}{\sqrt{-2it}}-2\sqrt{-2it}\frac{\partial}{\partial \bar
z})^N\cdot 1 \label{simpoprep}\end{equation} and we have operator
representation for our basic solution (\ref{hermitecomp})
\begin{equation}
\Psi_N(\bar z,t)=(\bar z+4it\frac{\partial}{\partial \bar z}
)^N\cdot1 \label{vorsolgentech}
\end{equation}
   This representation suggests the form of generating operator for solutions of
our equation (\ref{linearschrodinger}). By direct substitution we
can prove the next, \emph{vortex generation } technique for
solutions of this equation. If $\Phi(\bar z,t)$ is a solution of
(\ref{linearschrodinger}), then function
\begin{equation}
\Psi(\bar z,t)=(\bar z+4it\frac{\partial}{\partial \bar
z})\Phi(\bar z,t)
\end{equation}
is also a solution of (\ref{linearschrodinger}). This solution add
one vortex to the background configuration $\Phi(\bar z,t)$.
Applying it several times one can add arbitrary number of
vortices. Equation (\ref{linearschrodinger}) has evident solution
$\Phi=1$. Then
\begin{equation}
\Psi_1(\bar z,t)=(\bar z+4it\frac{\partial}{\partial \bar
z})\Phi(\bar z,t) =(\bar z+4it\frac{\partial}{\partial \bar
z})\cdot 1
\end{equation}
is also a solution of (\ref{linearschrodinger}). Next we have
\begin{equation}
\Psi_2(\bar z,t)=(\bar z+4it\frac{\partial}{\partial \bar
z})\Psi_1=(\bar z+4it\frac{\partial}{\partial \bar z})^2 \cdot1.
\end{equation}
Continuing this procedure we have solution of
(\ref{linearschrodinger}) for an arbitrary positive integer $N$
\begin{equation}
\Psi_N(\bar z,t)=(\bar z+4it\frac{\partial}{\partial \bar
z})\Psi_{N-1}=...=(\bar z+4it\frac{\partial}{\partial \bar z})^N
\cdot 1 .\label{basicN}
\end{equation}
This way we derived particular operator representation  for N
vortex solution \begin{equation} \Psi_N(\bar z,t)=
\prod_{i=1}^{N}(\bar z -{\bar z}_i (t)) =(\bar
z+4it\frac{\partial}{\partial \bar z})^N 1 .\end{equation} Using
previous result and linearity of (\ref{linearschrodinger}) we have
the next generalization. If $\Phi(\bar z,t)$ is a solution of
equation (\ref{linearschrodinger}) then function
\begin{equation}
\Psi(\bar z,t)= \sum_{n=0}^N a_n (\bar z+4it\frac{d}{d \bar z})^n
\,\Phi(\bar z,t)
\end{equation}
where $a_0,...,a_N$ are arbitrary constants, is also a solution of
equation (\ref{linearschrodinger}). As easy to see adding to the
system a new vortex in a proper way, we add an additional integral
of motion. Let us suppose that we have solution $\Psi (\bar z, t)$
 with $N$ - simple zeroes at points
$\bar z_1,...,\bar z_N$, that means, $\Psi (\bar z_n, t) = 0,
\,\,\,n = 1,...,N$. Then due to \begin{equation} \Psi (\bar z, t)
= \prod_{n=1}^{N} (\bar z - \bar z_n (t)) = \sum_{n=0}^N a_n
\Psi_n (\bar z, t)\end{equation} we have the system of N linear
algebraic equations $ \sum_{n=0}^N a_n \Psi_n (\bar z_k, t) = 0,
\,\,k = 1,...,N $. Extracting $n = 0$ term and dividing on $a_0$
it can be rewritten in the form of inhomogeneous system of N
algebraic equations \begin{equation} \sum_{n=1}^N b_n \Psi_n (\bar
z_k, t) = -1, \,\, k = 1,...,N\end{equation} on $N$ variables $b_n
= a_n/a_0$. Then, N integrals of motion can be found by Crammers
formulas $b_k = \Delta_k / \Delta, \,\, k = 1,...,N. $ From the
above consideration if we have $\Phi(\bar z,t)$ as
 a solution of equation
(\ref{linearschrodinger}) and $F(\bar z)$ is anti-analytic
function in some domain $D_0 = \{|z| < R \} $,  then function
\begin{equation}
\Psi(\bar z,t)= F(\bar z+4it\frac{d}{d \bar z}) \,\Phi(\bar z,t)
\end{equation}
 is also a solution of equation (\ref{linearschrodinger}).
 Since coefficients $a_n \equiv \frac{F^{(n)}(0)}{n !}$ are
integrals of motion, the function $F$ can be considered as the
generating function of integrals of motion. At the end of this
section we note that operator $K=\bar z+4it\partial_{\bar z}$ in
our approach is commuting with the Schr\"odinger
operator$S=i\partial_t+2\partial_{\bar z}^2$ and represent the
complex boost transformation of the Galilean group.
\section{Sine-Hermite Solution and Stationary Vortex Lattice}

As an application let us consider entire function \be F(w)= \sin w
= \sum_{n=0}^{\infty}(-1)^n \frac{w^{2n+1}}{(2n+1)!}.\ee Then
according to the above consideration \be \Phi(\bar z,t)=
\sum_{n=0}^{\infty} \frac{(-1)^n}{(2n+1)!}\left( \bar z+ 4it
\frac{\partial}{\partial \bar z} \right)^{2n+1}\cdot 1
\label{sinehermite}\ee is solution of equation
(\ref{linearschrodinger}). According to (\ref{vorsolgentech}) we
rewrite it as \be \Phi(\bar z,t)= \sum_{n=0}^{\infty}
\frac{(-1)^n}{(2n+1)!} \Psi_{2n+1}(\bar z,t)= \Psi_1 (\bar
z,t)-\frac{\Psi_3 (\bar z,t)}{3!}+ \frac{\Psi_5 (\bar
z,t)}{5!}-...\ee or by (\ref{hermitecomp})
\begin{eqnarray} \Phi(\bar z,t) = \sum_{n=0}^{\infty}
\frac{(-1)^n}{(2n+1)!}
(-2it)^{n+\frac{1}{2}}H_{2n+1}\left(\frac{\bar
z}{2\sqrt{-2it}}\right)
 = (-2it)^{n+\frac{1}{2}} H_1
\left(\frac{\bar z}{2\sqrt{-2it}}\right) \\
- \frac{1}{3!}(-2it)^{n+\frac{3}{2}} H_3 \left(\frac{\bar
z}{2\sqrt{-2it}}\right)+ \frac{1}{5!}(-2it)^{n+\frac{5}{2}} H_5
\left(\frac{\bar z}{2\sqrt{-2it}}\right)- ... \nonumber
\end{eqnarray} Writing operator
\begin{equation} \sin \left(\bar z+ 4it
\frac{\partial}{\partial \bar z} \right) = \frac{1}{2i}\left(\exp
\left[i\left( \bar z+4it \frac{\partial}{\partial \bar z}
\right)\right]- \exp \left[-i\left( \bar z+4it
\frac{\partial}{\partial \bar z} \right)\right]\right)
\end{equation} and using
Baker-Hausdorff relation for non-commuting operators $\bar z,
\frac{\partial}{\partial \bar z}$ we have
\begin{equation}
\sin \left(\bar z+ 4it \frac{\partial}{\partial \bar z} \right) =
- \frac{i}{2}e^{-2it}\left( e^{i\bar z}e^{-4t
\partial/\partial \bar z} - e^{-i\bar z}e^{4t
\partial/\partial \bar z}\right)
\end{equation}
or
\begin{equation}
\sin \left(\bar z+ 4it \frac{\partial}{\partial \bar z} \right) =
i\, e^{-2it}[\cos \bar z \,\sinh (4t \frac{\partial}{\partial \bar
z}) - i \sin \bar z \,\cosh (4t \frac{\partial}{\partial \bar z})]
\end{equation}
Applying this operator to evident solution $\Psi=1$ we get \be
\sin \left(\bar z+ 4it \frac{\partial}{\partial \bar z}
\right)\cdot 1= e^{-2it}\sin \bar z . \ee It describes the time
dependent  stationary vortex lattice solution  of
(\ref{linearschrodinger}). Positions of vortices in the lattice do
not change with time. Comparing with (\ref{sinehermite}) we have
the next expansion in terms of Hermite polynomials
\begin{eqnarray}
\exp (-2it) \sin \bar z = (-2it)^{n+\frac{1}{2}} H_1
\left(\frac{\bar z}{2\sqrt{-2it}}\right) \\-
\frac{1}{3!}(-2it)^{n+\frac{3}{2}} H_3 \left(\frac{\bar
z}{2\sqrt{-2it}}\right) + \frac{1}{5!}(-2it)^{n+\frac{5}{2}} H_5
\left(\frac{\bar z}{2\sqrt{-2it}}\right)- ...
\end{eqnarray}
 Then in terms of
$\Psi_n$ it gives \be \Phi(\bar z,t)=e^{-2it} \sin \bar z=
\Psi_1(\bar z,t)- \frac{\Psi_3(\bar z,t)}{3!}+\frac{\Psi_5(\bar
z,t)}{5!}-... \ee The last formula describe decomposition of the
stationary periodic lattice in terms of motion of odd
$1,3,5,...,2n+1,...$ number of vortices. It shows that proper
superposition of odd number basic vortex motions described in
Section 5 leads to the fixed in time periodic lattice of vortices
(vortex crystal).

\section{Single Vortex - Vortex Lattice Collision}
As another application of the results from Section 5, now we
construct a new class of solutions describing single vortex
collision with the vortex chain lattices. Consider solution of
(\ref{linearschrodinger}) in the double lattice form
\begin{equation}\zeta(\bar z ,t)= e^{-8it}\sin(\bar z-{\bar z}_1 (t))\sin(\bar
z+{\bar z}_1 (t)),\,\,\, \cos2{\bar z}_1(t)= r e^{8it}
\end{equation}
 then we have
another solution of (\ref{linearschrodinger}) in the form
\begin{eqnarray}
\Psi(\bar z , t)&=&(\bar z +4it\frac{\partial}{\partial \bar
z})\frac{1}{2}( r- e^{-8it}\cos 2\bar z)\nonumber \\
&=& \frac{1}{2}[ r e^{8it}\bar z- \bar z \cos 2\bar z+8it\sin
2\bar z]e^{-8it}
\end{eqnarray}
Using properties of trigonometric function of complex argument
\begin{equation} \sin 2\bar z = \sin 2x \cosh 2y - i \cos 2x \sinh 2y \end{equation} we
have for the real and imaginary parts of function $\Psi$ following
expressions correspondingly
$$
\Re \Psi = \frac{r}{2}x \cos 8t +\frac{r}{2}y \sin 8t + 8t \cos 2x
\sinh 2y- \frac{1}{2}x\cos 2x \cosh 2y - \frac{1}{2}y \sin 2x
\sinh 2y $$
$$
\Im \Psi = \frac{r}{2}x \sin 8t -\frac{r}{2}y \cos 8t+ 8t \sin 2x
\cosh 2y + \frac{1}{2}y\cos 2x \cosh 2y - \frac{1}{2}x \sin 2x
\sinh 2y $$ \newpage In figure 4 we show collision of a single
vortex with the double vortex lattices at positive time $t>0$. As
we can see, addition of the vortex leads to dimerization of the
lattice vortices which propagates in both directions and creates
finally the dimerized lattice.

\begin{figure}[h]
\begin{center} \epsfig{figure=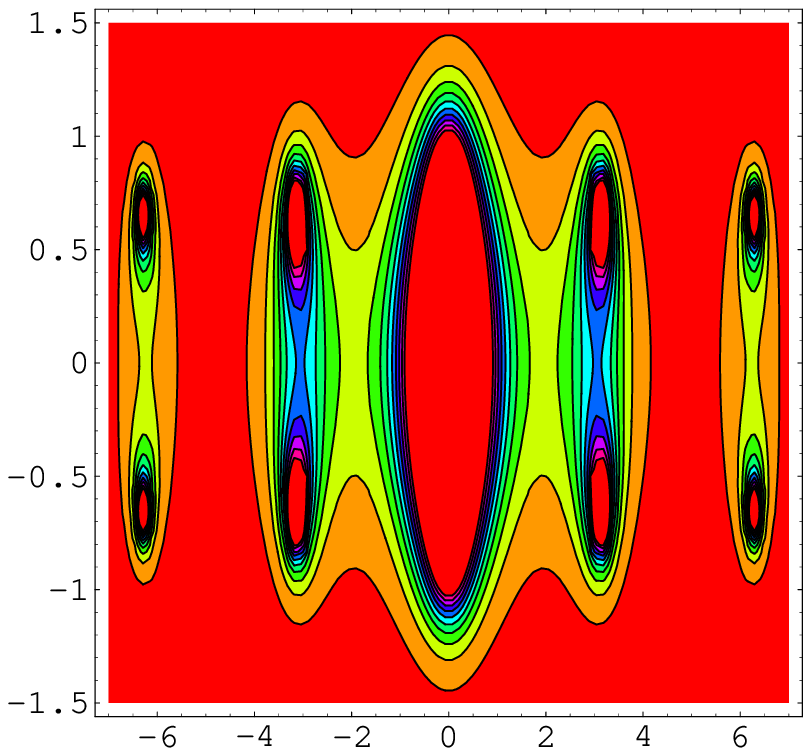,height=1.7cm,width=5cm}
 \epsfig{figure=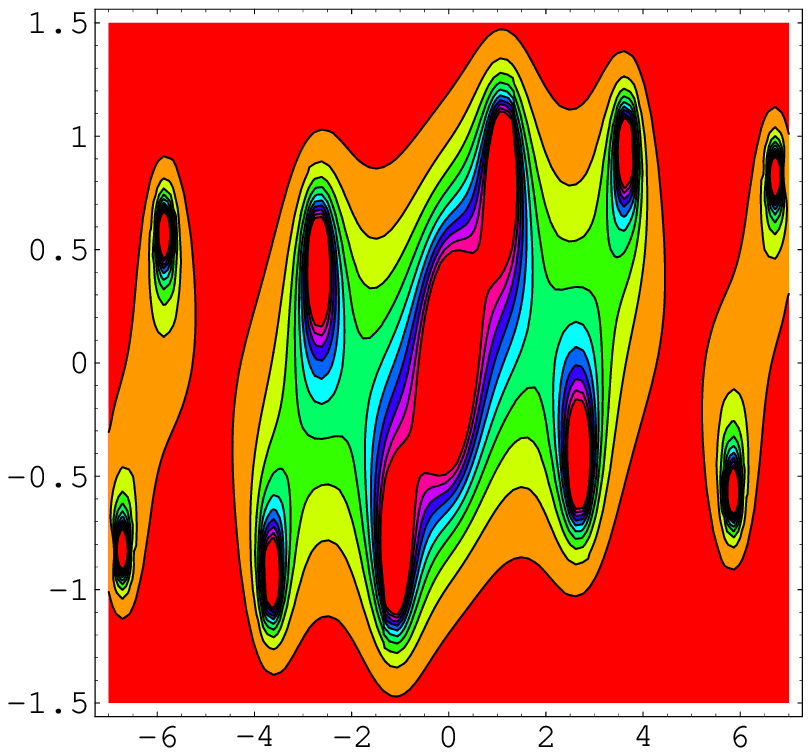,height=1.7cm,width=5cm}
 \epsfig{figure=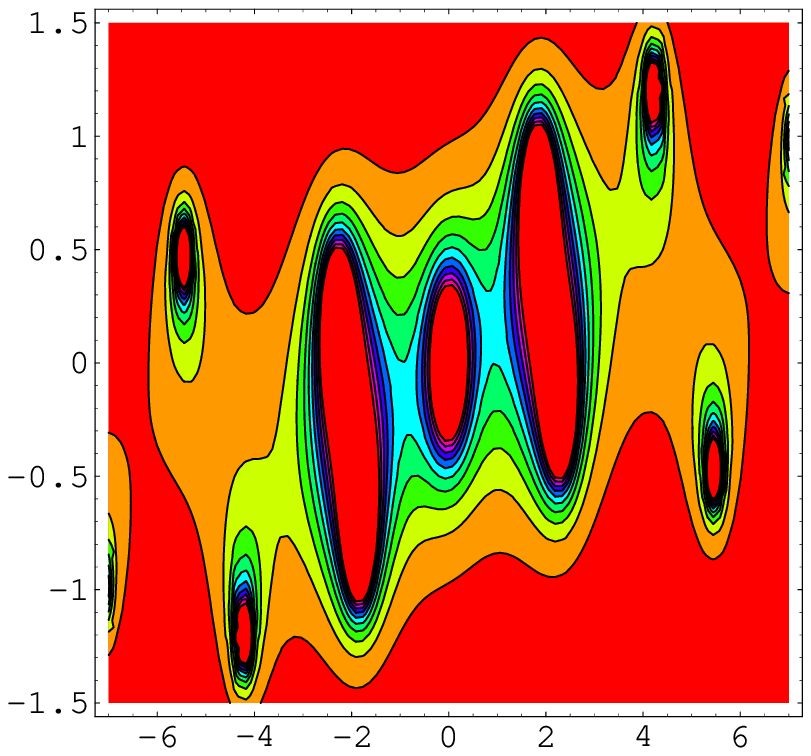,height=1.7cm,width=5cm}
 \epsfig{figure=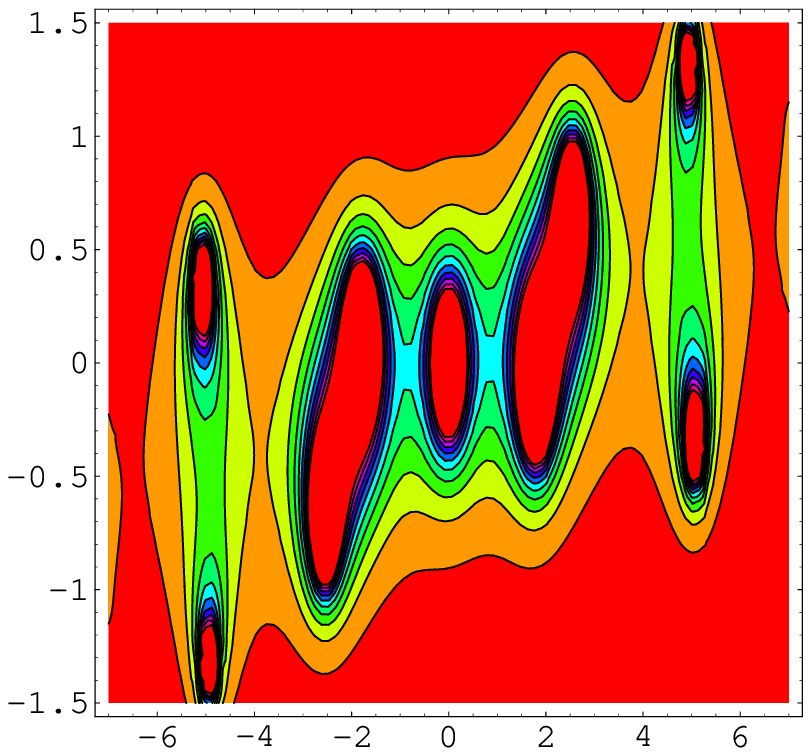,height=1.7cm,width=5cm}
\epsfig{figure=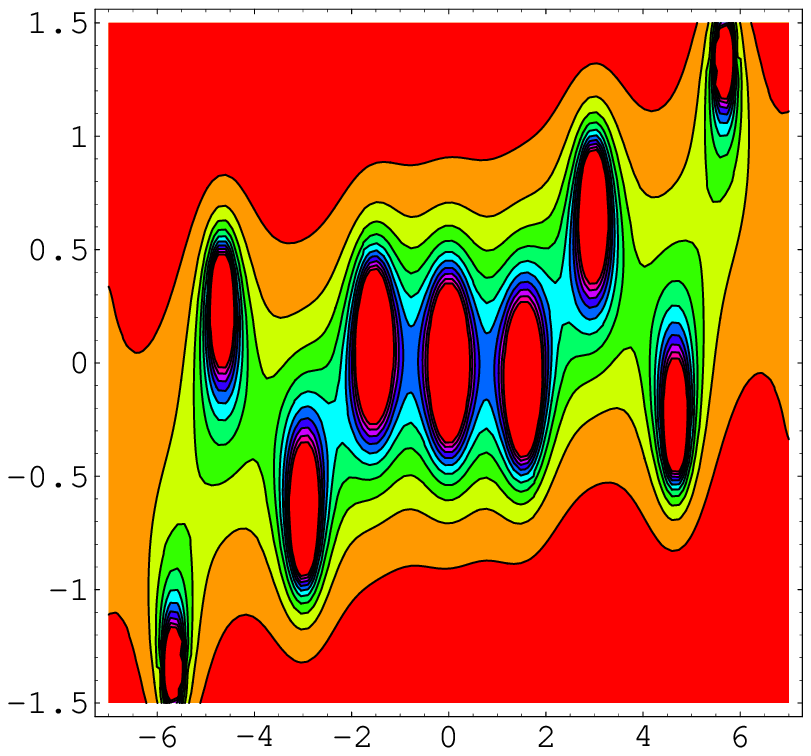,height=1.7cm,width=5cm}
\epsfig{figure=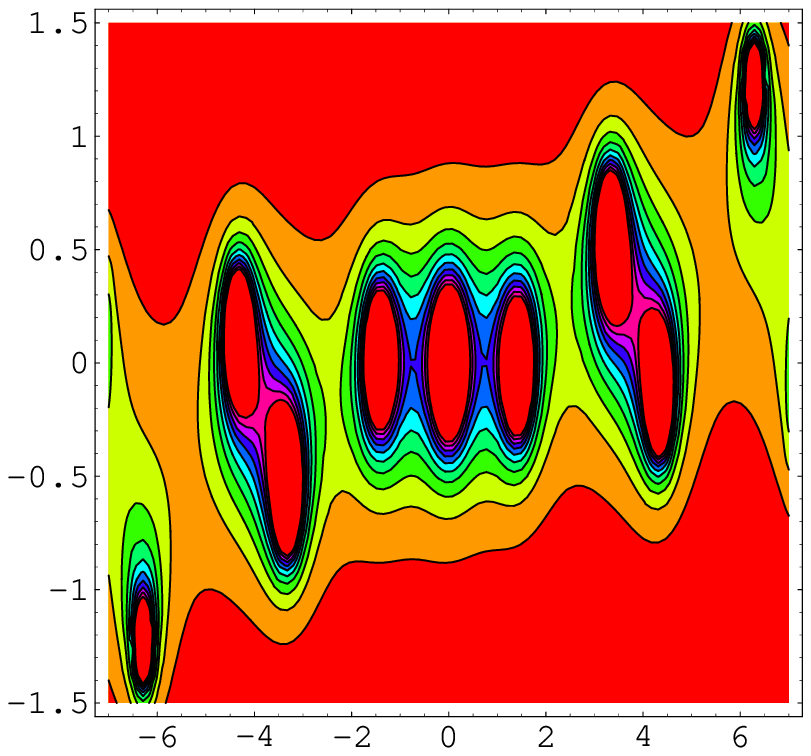,height=1.7cm,width=5cm}
\epsfig{figure=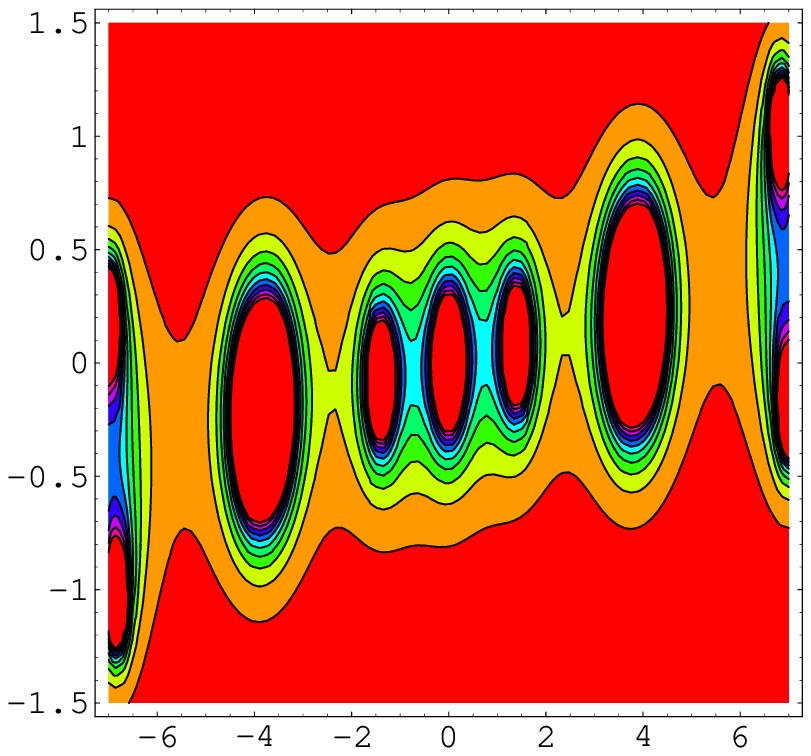,height=1.7cm,width=5cm}
\epsfig{figure=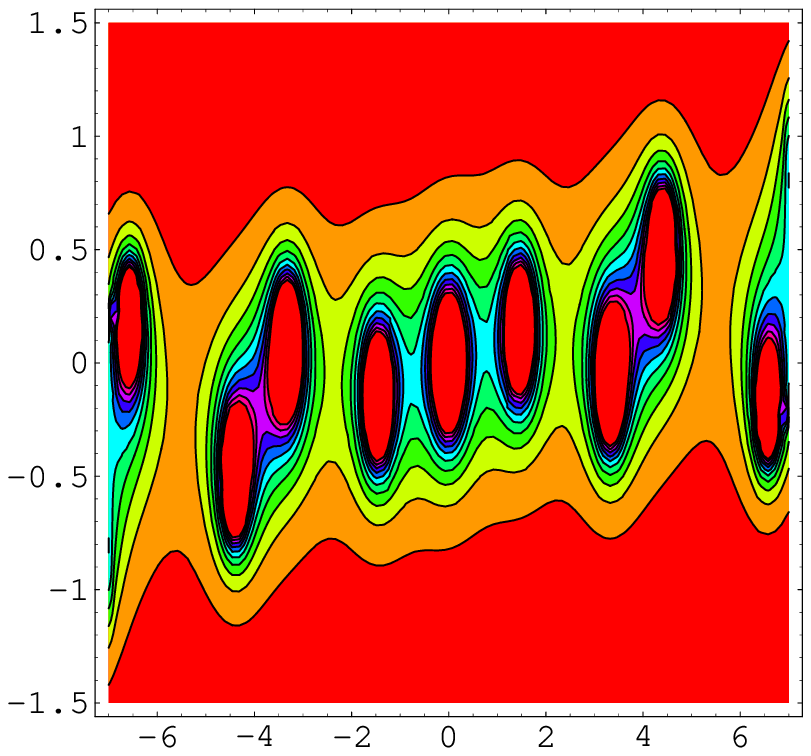,height=1.7cm,width=5cm}
\epsfig{figure=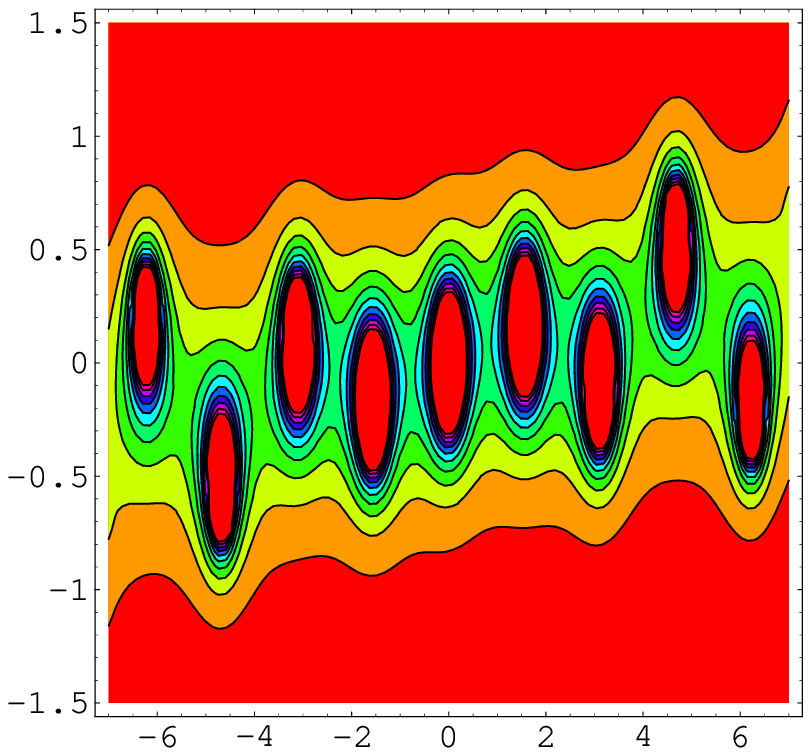,height=1.7cm,width=5cm}
\epsfig{figure=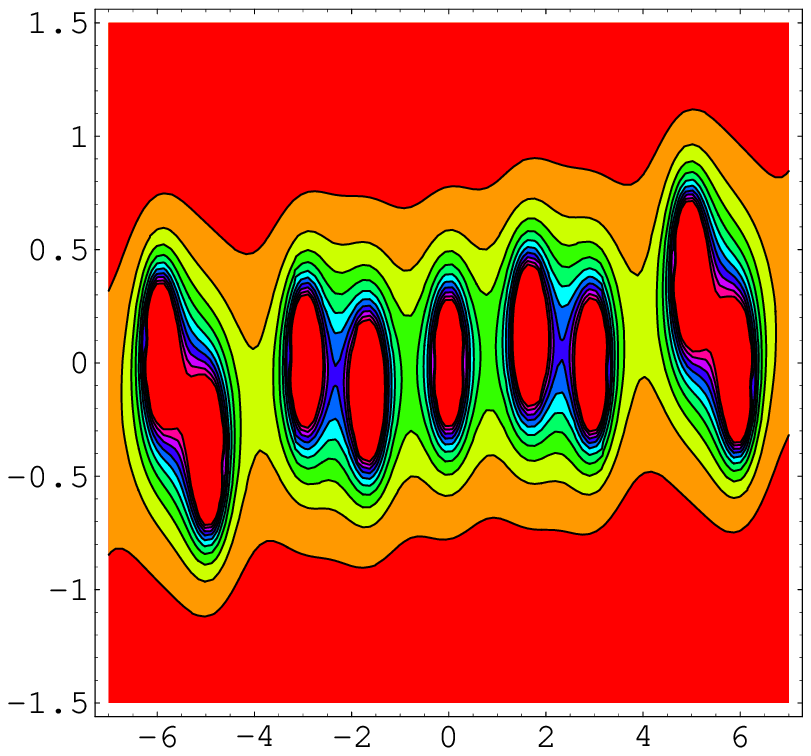,height=1.7cm,width=5cm}
\epsfig{figure=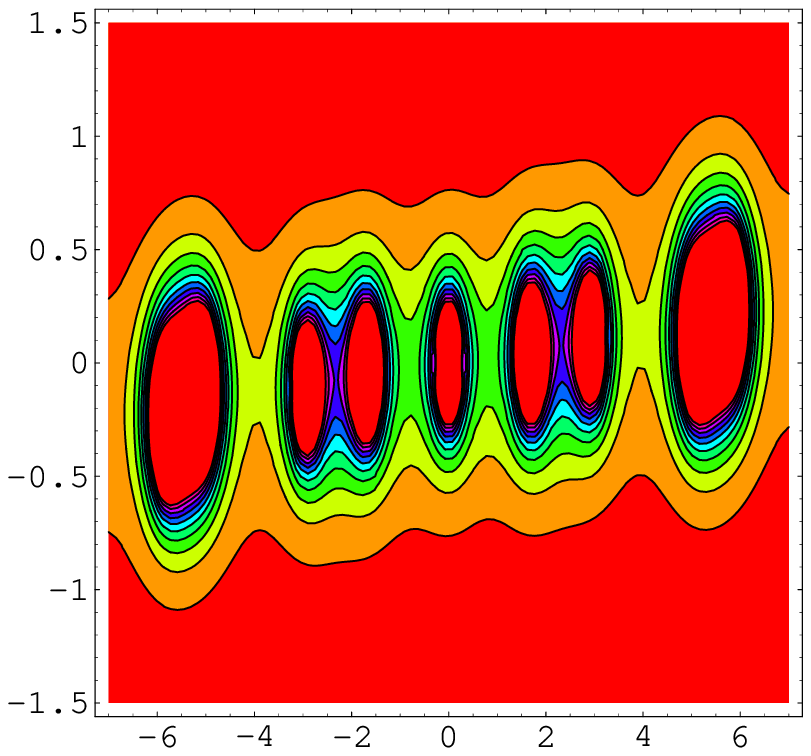,height=1.7cm,width=5cm}
\epsfig{figure=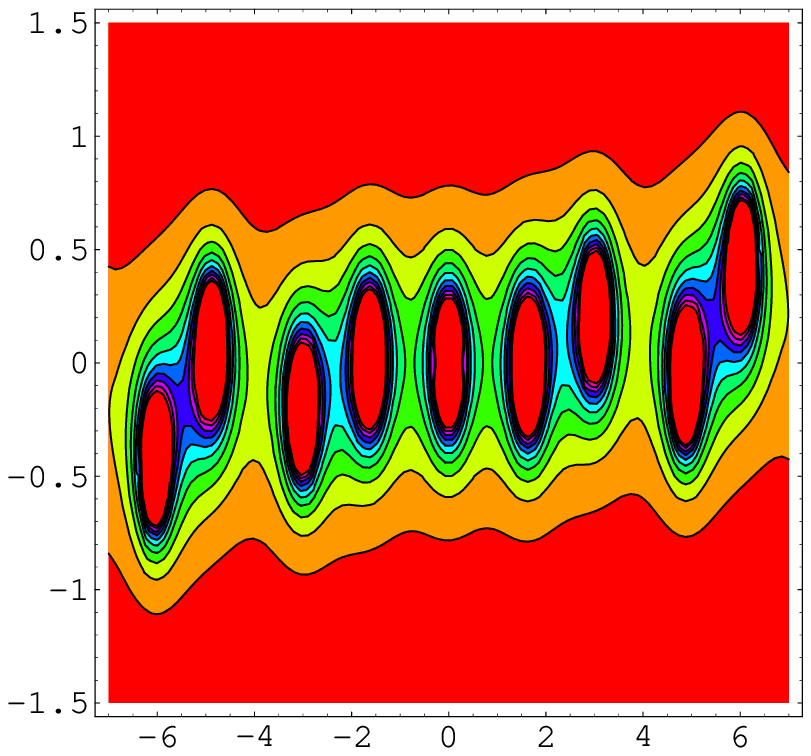,height=1.7cm,width=5cm}
\epsfig{figure=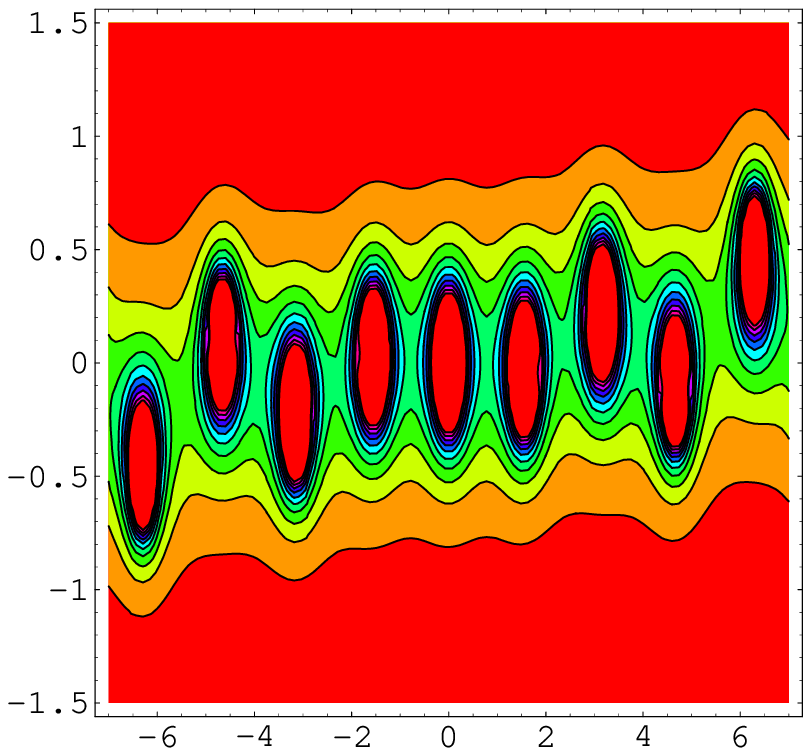,height=1.7cm,width=5cm}
\epsfig{figure=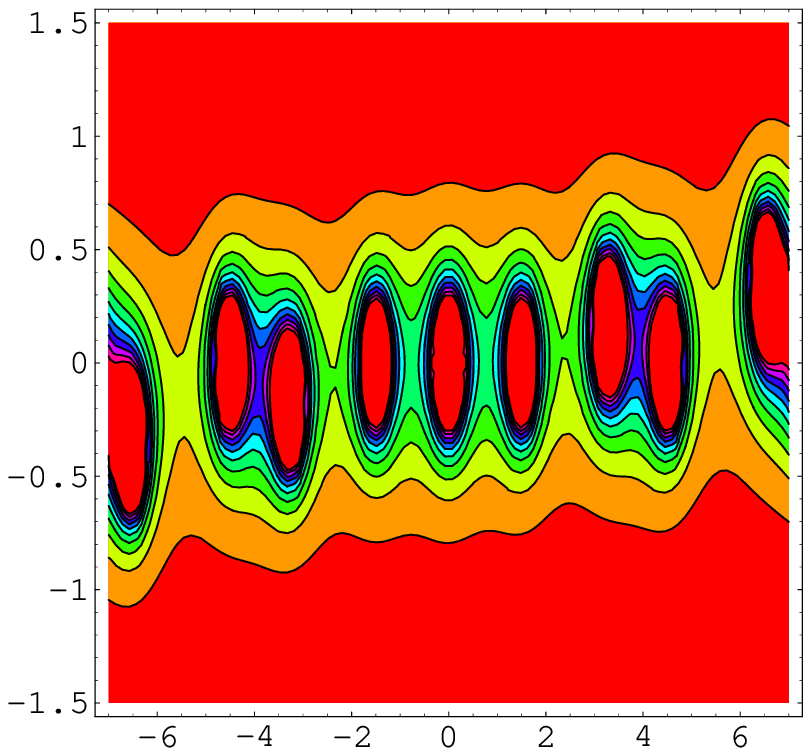,height=1.7cm,width=5cm}
\epsfig{figure=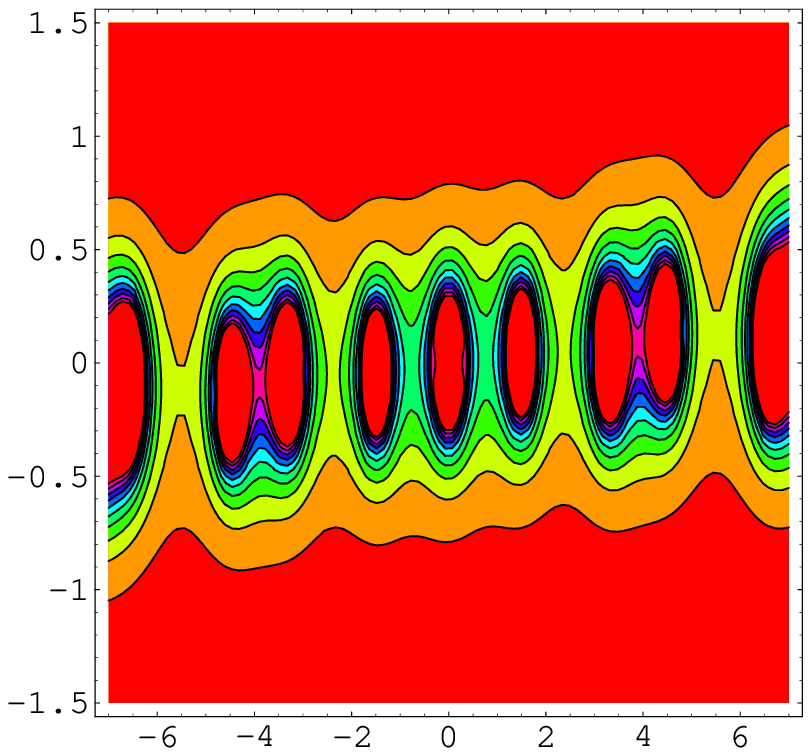,height=1.7cm,width=5cm}
\epsfig{figure=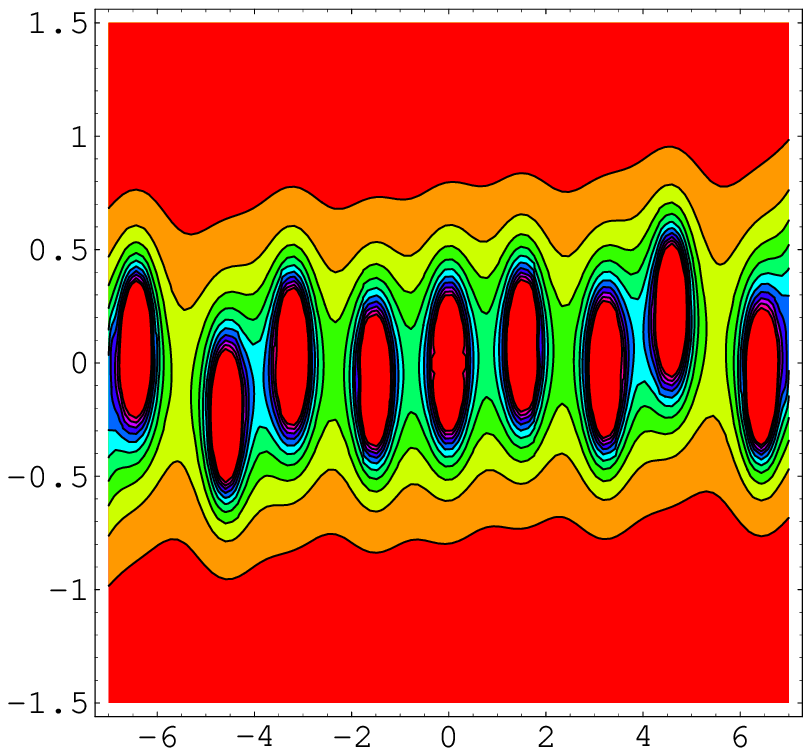,height=1.7cm,width=5cm}
\end{center}
\caption{Single Vortex - 2 Vortex Lattice Dynamics }
\label{V2L2-17}
\end{figure}

 Following the same idea we can consider solution
describing interaction of N- vortices with M- vortex chain
lattices (vortex crystal) in the form \begin{equation}\Psi(\bar z
, t)= e^{-2iM^2 t}(\bar z +4it\frac{\partial}{\partial \bar z})^N
\prod_{k = 1}^M \sin (\bar z - \bar z_k (t)).\end{equation}

\section{Integrable N-particle Problem for N-Vortex Motion}

In this section we show that problem of N-point vortices in the
plane can be reduced to complexified version of the Calogero-Moser
model type I \cite{Calogero}, \cite{Perelomov}. As we have seen in
section 4 the system of $N$ point vortices is described by
function (\ref{30}) satisfying the anti-holomorphic Schr\"odinger
equation (\ref{linearschrodinger}). Then positions of vortices in
the complex plane, $\bar{z}_1,...,\bar{z}_N$, are subject to the
first order system
\begin{equation}
\frac{d}{dt}\bar{z}_{j}=-4i\sum_{k\neq(j)}^N\frac{1}
{(\bar{z}_{j}-\bar{z}_{k})}\label{ishvortex1}.
\end{equation}
In one space dimension this system has been considered first in
\cite{Chood}, \cite{Calogero} for moving poles of Burgers'
equation, determined by zeroes of the heat equation. However,
complexification of the problem has several advantages. First of
all the root problem of algebraic equation degree N is complete in
the complex domain as well as the moving singularity analysis of
differential equations. In contrast to the one dimension, the pole
dynamics in the plane becomes time reversible (see discussion
after eq. (\ref{32})) and has interpretation of the vortex
dynamics.
 If we differentiate (\ref{ishvortex1}) once and use the system again then we  have
 the second order Newton's equations of
motion
\begin{equation}
\frac{d^{2}}{dt^{2}}\bar{z}_{j}=
\sum_{k\neq(j)}^N\frac{16}{(\bar{z}_{j}-\bar{z}_{k})^{3}}
\label{CaloMoser}.
\end{equation}
 These equations have Hamiltonian form
\begin{equation} \dot{\bar z}_j =
\frac{\partial H }{\partial p_j} = p_j,\,\,\, \dot p = -
\frac{\partial H}{\partial {\bar
z}_j}\label{compham}\end{equation} with Hamiltonian function
\begin{equation}
H=\frac{1}{2}\sum_{j=1}^{N}p^{2}_{j}+\sum_{j<k}\frac{8}{({\bar
z}_{j}- {\bar z}_{k})^2}. \label{H}\end{equation} The system
(\ref{CaloMoser}) is complexified version of the Calogero-Moser
system type I, where N-particle positions, $q_1,...,q_N$ should be
replaced by complex vortex positions ${\bar z}_1,...,{\bar z}_N$.
The Hamiltonian equations (\ref{compham}) are equivalent to the
Lax matrix equation
\begin{equation}
 i\dot L = A L- L A \label{ccmlax}
 \end{equation}
where
\begin{equation} L_{jk}
=\delta_{jk} p_ j+ig(1-\delta_{jk})\frac{1}{\bar z_j - \bar
z_k},\end{equation} \begin{equation} A_{jk} =g
\left[\delta_{jk}\sum_ {l\neq j}\frac{1}{(\bar z_j - \bar z_l)^2}
- (1-\delta_{jk})\frac{1}{(\bar z_j - \bar
z_k)^2}\right],\end{equation} and the coupling constant $g =
\sqrt{-4}$. Since matrix $L(t)$ is isospectrally deformed with
time, the corresponding (complex) eigenvalues are time independent
integrals of motion. Their symmetric functions as integrals of
motion, are given by
 $I_k = tr L^{k+1}$.
It shows that complexified Calogero-Moser system is an integrable
system and as a consequence, the N-vortex system
(\ref{ishvortex1}), which has been mapped to Calogero-Moser
system, is also integrable.

\section{ Integrable N-particle Problem for N-Vortex Crystals}

Similarly to the previous case now we consider mapping of the
N-vortex chain lattices (vortex crystals) to the complexified
Calogero-Moser system of type II and III \cite{Perelomov} . For
simplicity first we consider the system of two vortex chain
lattices described by function
\begin{equation} \zeta (\bar z, t) = e^{-4it}\sin (\bar z - \bar
z_1(t))\sin (\bar z - \bar
z_2(t))\label{vortexlattices1}\end{equation} so that position of
lattices is subject to the first order system \begin{equation}
\dot{\bar z_1}=-4 i\cot(\bar z_1 -\bar z_2) ,\,\,\, \dot{\bar
z_2}=4 i\cot(\bar z_1 -\bar z_2)\end{equation}
 Differentiating once  in time we get Newton's equations in the Hamiltonian form
\begin{equation} \dot{\bar z}_1 = \frac{\partial H}{\partial
p_1}=  p_1 ,\,\,\,\,\,\dot{ p}_1 = -\frac{\partial H}{\partial
\bar{z}_1}= 32 \frac{\cot
(\bar{z}_1-\bar{z}_2)}{\sin^3(\bar{z}_1-\bar{z}_2)}\end{equation}
\begin{equation}
\dot{\bar z}_2 = \frac{\partial H}{\partial p_2}= p_2
,\,\,\,\,\,\dot{p}_2 = -\frac{\partial H}{\partial \bar{z}_2}= 32
\frac{\cot (\bar{z}_2-\bar{z}_1)}{\sin^3(\bar{z}_2-\bar{z}_1)}
\end{equation}with Hamiltonian function
\begin{equation}
H=\frac{p_1 ^2 }{2}+\frac{p_2
^2}{2}+\frac{16}{\sin^2(\bar{z}_1-\bar{z}_2)}.
\end{equation}
Comparing this Hamiltonian of two vortex lattices with the
Calogero-Moser system, we realize that it corresponds to
complexified version of the model type III. We can generalize this
result considering $N$ vortex chain lattices in the horizontal
direction $x$. Positions of lattices are subject to the first
order system \begin{equation} \dot{\bar z_j}=-4 i\sum_{k=1,(k\neq
j)}^N \cot(\bar z_j -\bar z_k)\,\,\, j=1,...,N. \end{equation}
 Differentiating once we get
\begin{equation} \ddot{\bar z}_j
 = -32 \sum_{k=1,(k\neq
j)}^N  \frac{\cot
(\bar{z}_j-\bar{z}_k)}{\sin^2(\bar{z}_j-\bar{z}_k)}\,\,\,
j=1,...,N\end{equation} which is complexified Calogero-Moser
system type III with Hamiltonian
\begin{equation}
H=\frac{1}{2}\sum_j {p}_j ^2 + \sum_{j <
k}\frac{16}{\sin^2(\bar{z}_j-\bar{z}_k)}
\end{equation}
If instead of horizontal $x$ direction, we consider N chain
lattices in the vertical $y$ direction, it results in rotation of
every zero on angle $\pi/2$ and the replacement of complex
function $\sin \bar z$ by $\sinh \bar z$. As a result, the
corresponding Calagero-Moser system would be of type II. This
consideration shows also equivalence of complexified
Calogero-Moser systems of type II and III.
\section{ Time Dependent Schr\"odinger problem in Harmonic Potential}

Vorticity equation (\ref{17}) is invariant under substitution
$\psi \rightarrow \psi + U$   where $U $ is an arbitrary harmonic
function: $\Delta U = 0$. Choosing
\begin{equation}\psi=2\ln(1+|\zeta|^{2})+U(\bar{z},t)+\bar{U}(\bar{z},t)
\end{equation} and substituting to (\ref{zeta}) we have
complex Schr\"odinger equation with additional potential term
\begin{equation} i\zeta_{t}+ \zeta_{\bar z \bar z} +
\zeta_{\bar{z}}\bar{U}_{\bar{z}}=0
\label{antiholored3}\end{equation}

\section{Bound State of Vortices}

Here we choose particular form $ U(\bar z, t) = \frac{1}{2}\,\bar
z^2$ so that \begin{equation} \psi = 2\ln(1+|\zeta|^{2}) +
\frac{1}{2}(\bar z^2 + z).\end{equation} Then we have time
evolution subject to the equation \begin{equation}
i\zeta_{t}+2\zeta_{\bar{z}\bar{z}} + \bar z \zeta_{\bar z}=0.
\end{equation}
Looking for solution in the form \begin{equation} \zeta(\bar z, t)
= \sum_n e^{i n t} u_n (\bar z)\end{equation} we find that
function $u_n (\bar z)$ satisfies the complex Hermite equation
\begin{equation} u''_n + \bar z u'_n + u_n = 0.\end{equation} It gives time
dependent vortex solution in the form \be \zeta(\bar z, t) =
\sum_{n=0}^N e^{i n t} H_n (\bar z).\ee This solution is $N$-th
degree polynomial with periodic time dependent coefficients and it
describes the bound state of $N$ magnetic vortices. For particular
value $N = 2$ we have
\begin{equation} \zeta(\bar z, t)= H_0(\bar z)+e^{it} H_1(\bar z)+e
^{2it}H_2(\bar z),
\end{equation} \begin{equation} \Re \zeta = 1+2x \cos
t+ 2y\sin t+ [4(x^2- y^2)-2]\cos 2t+8xy \sin 2t \end{equation}
\begin{equation} \Im \zeta = -2y \cos t + 2x \sin t - 8xy \cos
2t+[4(x^2- y^2)-2]\sin 2t \end{equation}This solution is periodic
in time with period $T=2\pi$ and it describes the bound state of
two magnetic vortices (see Fig. 5).
\begin{figure*}[h]
\begin{center} \epsfig{figure=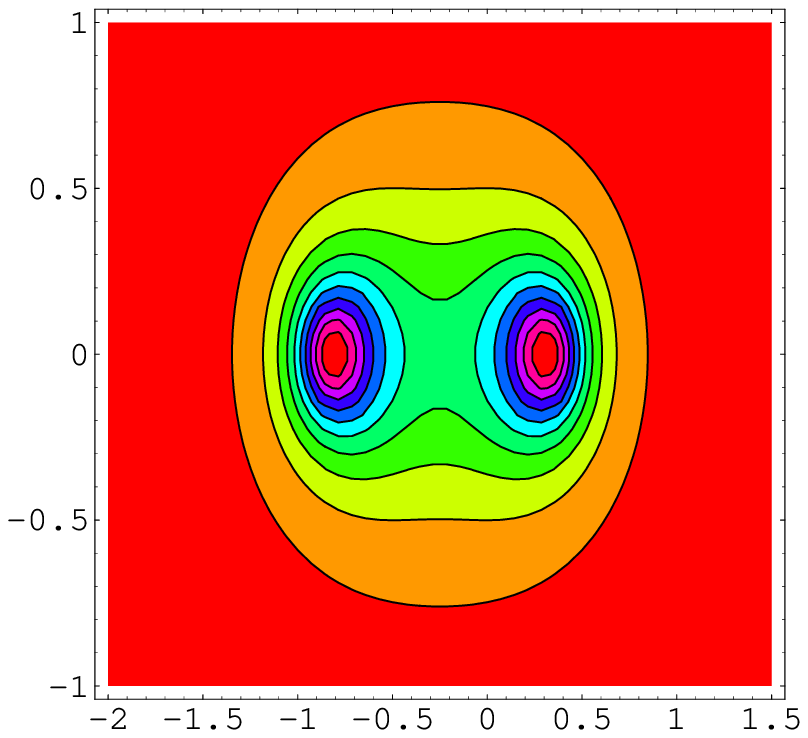,height=4cm,width=4cm}
\epsfig{figure=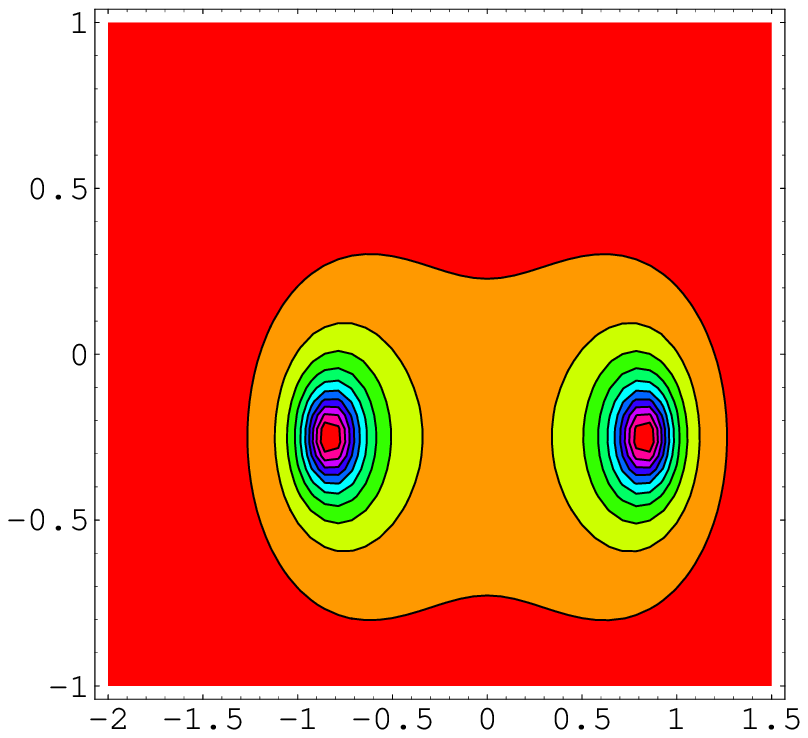,height=4cm,width=4cm}
\epsfig{figure=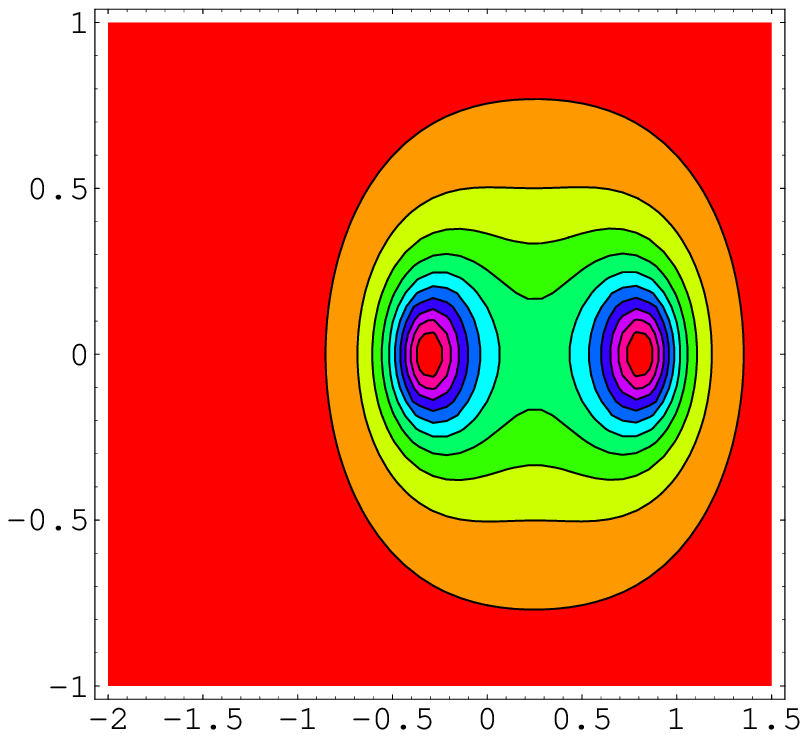,height=4cm,width=4cm}
\end{center}
\caption{Bound State of Two Magnetic Vortices} \label{Bound4}
\end{figure*}

\section{Static Vortex Crystals and the Liouville Equation}

From previous considerations we have seen that static vortex
configurations for Eq.(\ref{linearschrodinger}) are exist only for
N = 1. One can ask if there exist static N vortex or even N-vortex
lattices- vortex crystals for the Ishimori model
(\ref{ishmod1}),(\ref{ishmod2}). To answer this question we return
back to (\ref{zeta}). For static configurations $\partial
\zeta/\partial t = 0$ implies
\begin{equation}
\psi = 2\ln (1+|\zeta|^{2})-\ln\zeta_{\bar{z}} - \ln
\bar\zeta_{{z}} = -\ln \frac{|\zeta_{\bar
z}|^2}{(1+|\zeta|^{2})^2} \label{generalLiouville}\end{equation}
so that the vorticity equation (\ref{17}) is satisfied
automatically. But we notice that (\ref{generalLiouville}) is the
general solution of the Liouville equation \begin{equation} \Delta
\psi = 8 e^{-\psi}.\end{equation} If one chooses $
\zeta=\prod_{i=1}^{N}(\bar z- \bar z_i) $  with N simple zeroes in
the complex plane then it determines N static vortices located in
the plane at zeroes of this function \cite{Martina2}. To have
static vortex lattices periodic in $x$ direction we can consider
\begin{equation} \zeta=\prod_{i=1}^{N}\sin(\bar z- \bar z_i)
\end{equation}It determines N time independent magnetic vortex
lattices. For $N = 1$ the stream function \begin{equation} \psi =
\ln {\frac{(1+|\sin z|^2)^2}{|\cos z|^2}}\end{equation} describes
periodic in $x$ static lattice of vortices. If we choose
$\zeta=\bar z \sin{\bar z}$ then the stream function
\begin{equation} \psi= \ln {\frac{(1+|z|^2|\sin z|^2)^2}{|\sin
z+z\cos z|^2}}\end{equation} describes single vortex imposed on
vortex lattice.

%This solution describes the periodic lattice of vortices

\section{CONCLUSIONS}

\hskip 0.5cm To solve the problem of magnetic vortices in a planar
spin liquid model, first we found holomorphic reductions of the
model and showed that evolution equation at this reduction becomes
the linear anti-holomorphic Schr\"odinger equation. Analogy of the
anti-holomorphic Cole-Hopf transformation with the well known
hydrodynamical relation between the complex velocity and the
complex potential, leaded us to formulation of the complex
Burgers' equation with integrable $N$ vortex dynamics. We found
that vortices correspond to zeroes of the complex Schr\"odinger
equation. This allowed us to construct $N$ vortex configurations,
vortex chain lattices and their mutual dynamics in terms of the
complex Hermite polynomials. By mapping our vortex problem to
N-particle problem, the complexified Calogero-Moser system, we
showed its integrability and the Hamiltonian structure.

Finally we note that the holomorphic Hopf equation \be iu_t+uu_z=0
\ee which corresponds to the dispersionless limit of the
holomorphic Burgers' equation (\ref{u}), has been considered very
recently as nonlinear bosonisation in quantum hydrodynamics for
description of quantum shock waves in edge states of Fractional
Quantum Hall Effect \cite{Wiegmann}. The weak solution of this
equation for point vortices with strength $\Gamma_1,...,\Gamma_N$,
so that \be rot \,\, u = \sum_{k=1}^N \Gamma_k \delta(x-x_k(t))
\delta(y-y_k(t))\ee gives the following vortex system
\cite{pashaevg} \be \frac{dz_k}{dt}=4i\sum_{l=1,(l \neq k)}^N
\frac{\Gamma_l}{z_k-z_l},\,\,\, k=1,...,N \ee When all the vortex
strengths are equal $\Gamma_1=...=\Gamma_N$ then this system
reduces to (\ref{ishvortex1}) and is integrable. However, in the
general case the system is not known to be integrable. In
particular, for $N=3$ the system  with  constraint
$\Gamma_1=\Gamma_2\neq \Gamma_3$ has been studied in
\cite{CalSanSom} to explain the transition from regular to
irregular motion as travel on the Riemann surface.

\section{Acknowledgements}
This work was partially supported by TUBITAK under the Grant No.
106T447.

% ------------------------------------------------------------------------

\end{document}